\documentclass[12pt]{article}

\usepackage{natbib}
\usepackage{url} 
\usepackage{setspace}
\usepackage{xcolor}
\usepackage{bm}
\usepackage{amsthm,amsmath,amsfonts,amssymb,amsbsy,mathrsfs}
\usepackage[colorlinks,citecolor=blue,urlcolor=blue]{hyperref}
\usepackage{algorithm2e}
\usepackage{graphicx,psfrag,epsf}
\usepackage{enumerate}
\usepackage[toc,title,page]{appendix}
\usepackage{tabularx,booktabs}
\usepackage{placeins}
\usepackage{diagbox}
\usepackage{multirow}
\usepackage[createShortEnv]{proof-at-the-end}
\usepackage{comment}

\graphicspath{ {./images/} }

\newcommand{\blind}{1}
\theoremstyle{plain}

\newtheorem{theorem}{Theorem}[section]

\theoremstyle{remark}
\newtheorem{definition}[theorem]{Definition}

\newcommand\setrow[1]{\gdef\rowmac{#1}#1\ignorespaces}
\newcommand\clearrow{\global\let\rowmac\relax}
\newcolumntype{Y}{>{\centering\arraybackslash}X}

\pgfkeys{/prAtEnd/local custom defaults/.style={
 text link={See \hyperref[proof:prAtEnd\pratendcountercurrent]{proof} in Appendix \ref{apdxAproof}.}
}
}

\DeclareMathOperator*{\argmax}{arg\,max}  
\DeclareMathOperator*{\argmin}{arg\,min}  

\makeatletter
\renewcommand\section{\@startsection {section}{1}{\z@}%
                                   {-3.5ex \@plus -1ex \@minus -.2ex}%
                                   {2.3ex \@plus.2ex}%
                                   {\normalfont\fontfamily{phv}\fontsize{13}{16}\bfseries}}
\renewcommand\subsection{\@startsection{subsection}{2}{\z@}%
                                     {-3.25ex\@plus -1ex \@minus -.2ex}%
                                     {1.5ex \@plus .2ex}%
                                     {\normalfont\fontfamily{phv}\fontsize{11}{14}\bfseries}}
\renewcommand\subsubsection{\@startsection{subsubsection}{3}{\z@}%
                                    {-3.25ex\@plus -1ex \@minus -.2ex}%
                                     {1.5ex \@plus .2ex}%
                                     {\normalfont\normalsize\fontfamily{phv}\fontsize{11}{14}\selectfont}}

\begin{document}

\def\spacingset#1{\renewcommand{\baselinestretch}%
{#1}\small\normalsize} \spacingset{1}


\if1\blind
{
	\title{\bf {Overlapping Indices for Dynamic Information Borrowing in Bayesian Hierarchical Modeling}}
	\author{ Xuetao Lu $^a$ and J. Jack Lee $^{a}$\thanks{\small {Author for correspondence: jjlee@mdanderson.org}}\\
	$^a$ \small {Department of Biostatistics, The University of Texas MD Anderson Cancer Center} }
	\date{}
  \maketitle
} \fi

\if0\blind
{
  \bigskip
  \bigskip
  \bigskip
  \begin{center}
    {\LARGE\bf Title}
\end{center}
  \medskip
} \fi

\bigskip
\begin{abstract}
Bayesian hierarchical model (BHM) has been widely used in synthesizing information across subgroups. Identifying heterogeneity in the data and determining proper strength of borrow have long been central goals pursued by researchers. Because these two goals are interconnected, we must consider them together. This joint consideration presents two fundamental challenges: (1) How can we balance the trade-off between homogeneity within the cluster and information gain through borrowing? (2) How can we determine the borrowing strength dynamically in different clusters? To tackle challenges, first, we develop a theoretical framework for heterogeneity identification and dynamic information borrowing in BHM. Then, we propose two novel overlapping indices: the overlapping clustering index (OCI) for identifying the optimal clustering result and the overlapping borrowing index (OBI) for assigning proper borrowing strength to clusters. By incorporating these indices , we develop a new method BHMOI (Bayesian hierarchical model with overlapping indices). BHMOI includes a novel weighted K-Means clustering algorithm by maximizing OCI to obtain optimal clustering results, and embedding OBI into BHM for dynamically borrowing  within clusters. BHMOI can achieve efficient and robust information borrowing with desirable properties. Examples and simulation studies are provided to demonstrate the effectiveness of BHMOI in heterogeneity identification and dynamic information borrowing.
\end{abstract}

\noindent%
\vfill

\newpage
\spacingset{1.9} 
\section{Introduction}
\label{sec:intro}
Information borrowing has emerged as a promising approach in data analysis involving multiple sources or subgroups.  One prime example is analyzing data from clinical trials. When external data exists, information borrowing leverages it to supplement the current trial data. Several methods, such as power prior \citep{Ibrahim1998}, meta-analytic-predictive (MAP) priors \citep{Neuenschwander2010}, and commensurate prior \citep{Hobbs2011}, have been proposed to control the amount of information borrowed by accounting for the similarity between the external and the trial data. In other cases, external data may not be available, and but the trial data may involve multiple subgroups, such as in basket trials \citep{Park2019} or multi-regional studies. In these situations, a Bayesian hierarchical model (BHM) can allow information borrowing across subgroups rather than analyzing each subgroup's data separately or lump data from all subgroups together. With appropriate priors, BHM can improve parameter estimation accuracy and lower type I error or increase power \citep{berry2010}. However, the presence of heterogeneity among subgroups can hinder the benefits of BHM \citep{Freidlin2013}. In such situations, one may further form subgroups into clusters such that subgroups within clusters are more homogeneous and between clusters are more heterogeneous. The crucial yet challenging tasks are to determine how to form the clusters and how to select suitable priors to achieve efficient and robust dynamic information borrowing.

To accomplish these tasks, two interconnected issues need to be addressed: identifying heterogeneity among subgroups to form clusters and determining the amount of information to borrow based on the homogeneity within each cluster. These issues should be considered together since the amount of information borrowed must be quantified based on the homogeneity of subgroups in clusters, which is the core idea behind dynamic information borrowing. Two challenges arise from these issues. Firstly, we need to balance the trade-off between within-cluster homogeneity and the efficiency gain through information borrowing. Clustering is typically employed to identify heterogeneity, then forming homogeneous clusters. Accurate clustering usually requires a higher proportion of homogeneous elements, resulting in smaller clusters, i.e., fewer elements in the cluster. However, efficiency gain and robust inference usually require larger clusters. But as the size of the cluster increases, the heterogeneity also increases which limits the amount of information borrowing. These are two critical elements yet with competing goals in information borrowing \citep{Schmidli2014,Neuenschwander2016,Zhang2021}. Increasing the homogeneity of clusters and keeping the efficiency and robustness of information borrowing within each cluster should be well-balanced. Secondly, we should dynamically quantify borrowing strength in different clusters, which has been one of the central issues of dynamic information borrowing \citep{Nikolakopoulos2018,Chu2018}. Existing methods \citep{LenNovelo2012,Hobbs2018,Chen2019,Chen2020,Yin2021,Kang2021} did not adequately consider the trade-off between within-cluster homogeneity and the efficiency gain through information borrowing. Additionally, they typically treated the subgroup clustering and information borrowing separately, employing a fixed common prior to control borrowing strength across all clusters, regardless of any homogeneity differences that may exist among them.

To address the aforementioned issues and challenges, we introduce a BHM framework for heterogeneity identification and dynamic information borrowing. In addition, we propose two overlapping indices, the overlapping clustering index (OCI) and the overlapping borrowing index (OBI), which are constructed based on the concept of overlapping coefficient. OCI aids in heterogeneity identification through clustering, while OBI determines the borrowing strength for each cluster. \cite{Xu2020} also proposed an individual borrowing strength measure (InBS) and an overall borrowing index (OvBI) based on Mallow’s distance \citep{Mallows1972} to characterize the strength of borrowing behaviors among subgroups in BHM. However, InBS and OvBI primarily serve as exploratory tools for understanding the roles of priors, rather than determining appropriate borrowing strength for heterogeneous subgroups. By incorporating OCI and OBI into the framework, we propose a novel method BHMOI (Bayesian hierarchical model with overlapping indices), which comprises an OCI-guided weighted K-Means clustering algorithm and a method for embedding OBI into the BHM to determine the proper cluster size and the amount of borrowing within each cluster.  Thus, BHMOI can strike a balance between obtaining efficiency while maintaining robustness for statistical inference on the parameter of interest under BHM.

In Section \ref{sec:sec2}, we present the framework for heterogeneity identification and dynamic information borrowing in BHM. The construction of the two overlapping indices is discussed in Section \ref{sec:sec3}. In Section \ref{sec:sec4}, we detail the development of BHMOI. To compare the performance of BHMOI with existing methods, we conduct simulation and real data studies in Section \ref{sec:sec5}. Finally, we conclude the paper with a brief discussion in Section \ref{sec:Con}. Appendix \ref{apdxA} contains proofs and examples, and an open-source R implementation of BHMOI is available on GitHub at https://github.com/lxtpvt/BHMOI.git.

\section{Heterogeneity Identification and Dynamic Information Borrowing in BHM}
\label{sec:sec2}
\subsection{Problem and Assumption}

We consider a set of observations in sample space $\mathcal{Y}$ denoted as $\boldsymbol{y} = \{y_1, \dots, y_n\}$, with corresponding parameters in the space $\Omega$ denoted as $\boldsymbol{\theta} = \{\theta_1, \dots, \theta_n\}$. The Bayesian model for the posterior of $\boldsymbol{\theta}$ can be represented in \eqref{eq:BHM-0}, which is widely used in the study of medicine, including clinical trials and drug development \citep{Lee2012, Tidwell2019}.
\begin{equation}\label{eq:BHM-0}
     p(\boldsymbol{\theta} \mid \boldsymbol{y})={\frac {p(\boldsymbol{y}\mid \boldsymbol{\theta} )p(\boldsymbol{\theta} )}{p(\boldsymbol{y})}}
\end{equation}
Traditionally, the assumption of exchangeability is made for ${\theta_1, \dots, \theta_n}$. In this paper, we relax this assumption and consider the possibility of non-exchangeability among them. Based on the concept of exchangeability, we define homogeneity and heterogeneity as follows, which will be frequently used throughout this paper.
\begin{definition}[Homogeneity versus heterogeneity]\label{df:heter}
Let $S$ be a map defined in \eqref{partitionMap} that partitions $\{\theta_1, \dots, \theta_n\}$ into $K$ mutually exclusive clusters, where $n_j$ is the number of parameters in cluster $j$.
\begin{equation}\label{partitionMap}
S: \hspace{2mm} \{\theta_1, \dots, \theta_n\} \mapsto \{ \boldsymbol{\theta_1}, \dots, \boldsymbol{\theta_K} \} =\{\{\theta_{11},\dots ,\theta_{1n_1}\}, \dots, \{\theta_{K1},\dots, \theta_{Kn_K}\}\}
\end{equation}
If the parameters within each cluster $\boldsymbol{\theta_j}=\{\theta_{j1},\dots, \theta_{jn_j}\}$ are exchangeable but not between clusters, we say that the parameters are homogeneous within clusters and heterogeneous between clusters.
\end{definition}

\subsection{Dynamic Information Borrowing Within Each Cluster}\label{sec:DIBaHC}
Dynamic information borrowing aims to determine the amount of information to borrow within a cluster based on the degree of homogeneity of the elements within it. The more homogeneous the cluster, the more information can and should be borrowed. Suppose $\{\theta_1, \dots, \theta_n\}$ can be partitioned into $K$ heterogeneous clusters by $S$ in \eqref{partitionMap}. To implement dynamic information borrowing, we partition the model \eqref{eq:BHM-0} into $K$ heterogeneous clusters, resulting in the model:
\begin{equation}\label{eq:BHM-2}
     p(\boldsymbol{\theta_j} \mid \boldsymbol{y_j})={\frac {p(\boldsymbol{y_j}\mid \boldsymbol{\theta_j} )p(\boldsymbol{\theta_j} )}{p(\boldsymbol{y_j})}} =\frac {p(\boldsymbol{y_j}\mid \boldsymbol{\theta_j} )}{p(\boldsymbol{y_j})} \int p(\boldsymbol{\theta_j} \mid \boldsymbol{\eta_j} ) p(\boldsymbol{\eta_j}) \,d \boldsymbol{\eta_j}, \hspace{3mm} j=1,\cdots,K.
\end{equation}
Here, $\boldsymbol{\eta}$ are the hyperparameters for model specification. We can partition $\boldsymbol{\eta}$ into three subsets, $\boldsymbol{\eta}=\{\boldsymbol{\eta_{c}},\boldsymbol{\eta_{b}},\boldsymbol{\eta_{o}}\}$: $\boldsymbol{\eta_{c}}$ represents the hyperparameters related to clustering, $\boldsymbol{\eta_{b}}$ represents those associated with information borrowing strength, and $\boldsymbol{\eta_{o}}$ includes all additional hyperparameters for model specification. Instead of assuming full independence, $p(\boldsymbol{\eta_{c}},\boldsymbol{\eta_{b}},\boldsymbol{\eta_{o}}) = p(\boldsymbol{\eta_{c}})p(\boldsymbol{\eta_{b}})p(\boldsymbol{\eta_{o}})$, we adopt partial dependence, $p(\boldsymbol{\eta_{c}},\boldsymbol{\eta_{b}},\boldsymbol{\eta_{o}}) = p(\boldsymbol{\eta_{o}})p(\boldsymbol{\eta_{b}} | \boldsymbol{\eta_{c}})p(\boldsymbol{\eta_{c}})$, because the information borrowing strength depends on the clustering result. Then, for each cluster $\boldsymbol{\theta_j}=\{\theta_{j1},\dots, \theta_{jn_j}\}$, we have the model:
\begin{equation}\label{eq:BHM}
     p(\boldsymbol{\theta_j} \mid \boldsymbol{y_j}) = \frac {p(\boldsymbol{y_j}\mid \boldsymbol{\theta_j} )}{p(\boldsymbol{y_j})} \int p(\boldsymbol{\theta_j} \mid \boldsymbol{\eta_{jc}},\boldsymbol{\eta_{jb}},\boldsymbol{\eta_{jo}} ) p(\boldsymbol{\eta_{jo}}) p(\boldsymbol{\eta_{jb}}|\boldsymbol{\eta_{jc}})p(\boldsymbol{\eta_{jc}}) \,d \boldsymbol{\eta_j}
\end{equation}
Clusters may have varying degrees of homogeneity, and we achieve dynamic information borrowing by assigning different borrowing strengths across clusters according to their degree of homogeneity. The procedure includes two steps:
\begin{itemize}
    \item[(a)] Determine $p(\boldsymbol{\eta_{jc}})$.
    \item[(b)] Determine $p(\boldsymbol{\eta_{jb}}|\boldsymbol{\eta_{jc}})$ according to the degree of homogeneity of the clusters.
\end{itemize}
The method for Determining $p(\boldsymbol{\eta_{jc}})$ will be discussed in Section \ref{sec:HI}, while the key to measuring the degree of homogeneity of distributions in a cluster will be discussed in Section \ref{sec:obi}.

\subsection{Heterogeneity Identification for Cluster Partition}\label{sec:HI}
The purpose of heterogeneity identification is to determine the partition $S$ in \eqref{partitionMap}, which is a discrete random variable with a probability mass function $p(S)$. The hyperparameters $\boldsymbol{\eta_{c}}$ are only related to clustering, and thus $p(S)=p(\boldsymbol{\eta_{c}})$. Therefore, the task of determining $p(\boldsymbol{\eta_{c}})$ in Section \ref{sec:DIBaHC} can be reformulated as finding $p(S)$, which can be approximated using the following expression:
\begin{equation}\label{eq:empBayes-1}
p(S) \approx \tilde{p}(S|\boldsymbol{y}) = \int p(S|\boldsymbol{\theta})\tilde{p}(\boldsymbol{\theta}|\boldsymbol{y}) \,d \boldsymbol{\theta}
\end{equation}
Here, $\tilde{p}(\boldsymbol{\theta}|\boldsymbol{y})=p(\boldsymbol{y}|\boldsymbol{\theta})\tilde{p}(\boldsymbol{\theta})$, where $\tilde{p}(\boldsymbol{\theta})$ represents a non-informative prior of $\boldsymbol{\theta}$.

From equation \eqref{eq:empBayes-1}, we know that $\tilde{p}(S|\boldsymbol{y})$ is determined by $p(S|\boldsymbol{\theta})$ and $\tilde{p}(\boldsymbol{\theta}|\boldsymbol{y})$. The latter is determined by the data $\boldsymbol{y}$, while the former is determined by the algorithm of distribution clustering that creates the partition $S$ with given $\tilde{p}(\boldsymbol{\theta}|\boldsymbol{y})$. In practice, when we fix the data $\boldsymbol{y}$ and the distribution clustering algorithm (see Section \ref{sec:PMDC}), $\tilde{p}(S|\boldsymbol{y})$ will concentrate to a point.
\begin{equation}\label{eq:empBayes-2}
\tilde{p}(S^*|\boldsymbol{y})\approx 1, \hspace{3mm} S^*=\argmax_{S}\tilde{p}(S|\boldsymbol{y})
\end{equation}
Because of $p(S)=p(\boldsymbol{\eta_{c}})$, we can obtain that $p(\boldsymbol{\eta_{jb}}|\boldsymbol{\eta_{jc}})p(\boldsymbol{\eta_{jc}})\approx p(\boldsymbol{\eta_{jb}}|\boldsymbol{\eta^*_{jc}})=p(\boldsymbol{\eta_{jb}}|S^*)$. Thus, model \eqref{eq:BHM} can be converted to model \eqref{eq:BHM-F} as follows.
\begin{equation}\label{eq:BHM-F}
     p(\boldsymbol{\theta_j} \mid \boldsymbol{y_j}) \approx \frac {p(\boldsymbol{y_j}\mid \boldsymbol{\theta_j} )}{p(\boldsymbol{y_j})} \int p(\boldsymbol{\theta_j} \mid \boldsymbol{\eta_{jb}},\boldsymbol{\eta_{jo}} ) p(\boldsymbol{\eta_{jo}}) p(\boldsymbol{\eta_{jb}}|S^*) \,d \boldsymbol{\eta_{jo}} \,d \boldsymbol{\eta_{jb}}
\end{equation}

\subsection{Optimal Partition Through Distribution Clustering}\label{sec:PMDC}
We apply the distribution clustering algorithm to find the optimal partition $S^*$ in \eqref{eq:empBayes-2}. There are two main branches of study in distribution clustering: one based on the K-Means algorithm \citep{Nielsen2014, Krishna2019}, and the other on nonparametric Bayesian methods using priors such as the Dirichlet process \citep{Chen2020}. Note that, the nonparametric Bayesian-based approach is not suitable for our framework due to two reasons. First, the hyperparameters controlling clustering in the prior are difficult to specify, as mentioned in Section \ref{sec:simDesign}. Second, the clustering and information borrowing steps are independent and disjoint, which is not ideal. Most commonly used clustering methods have no explicit homogeneity measure in the clustering step that can be used to determine the borrowing strength in the information borrowing step. To address these issues, we choose the K-Means algorithm and make additional improvements. Let $(\mathcal{A},\mathscr{F},P)$ be a probability space with $\mathcal{A}$ as the sample space, $\mathscr{F}$ as the $\sigma$-algebra of $\mathcal{A}$, and $P$ as a probability measure on $\mathscr{F}$. We consider the space of all random variables with probability density or mass functions in the probability space, denoted by $\mathcal{M}$. The K-Means algorithm is conducted in the metric space $(\mathcal{M},d)$, where the distance $d$ is defined as the total variation distance between two random variables $\xi$ and $\zeta$ with probability density or mass functions $f$ and $g$ respectively.
\begin{equation}\label{eq:dtv}
     d(\xi,\zeta) = \sup|\xi-\zeta| =
     \begin{cases}
        1-\int_\Omega min\{f(t),g(t)\}\,dt, & \text{continuous distributions}\\
        1-\sum_\Omega min\{f(t_i),g(t_i)\}I(t_i \in \Omega), & \text{discrete distributions}
    \end{cases}
\end{equation} 
where $\Omega$ is the common support of $f$ and $g$. Given a set of random variables $\{\xi_1, \ldots, \xi_n\}$ in $\mathcal{M}$, and let $S=\{S_1,\ldots,S_K\}$ denote the $K$ partitions of them, where $1\leq K\leq n$. The K-Means algorithm aims to find the optimal partition $S^*$ in \eqref{eq:empBayes-2} by minimizing the sum of distances between each random variable and its assigned cluster centroid.
\begin{equation}\label{eq:km}
S^{*} = \argmin_{S} \sum^K_{m=1}\sum^n_{i=1} d(\xi_i,\zeta_m) \cdot I(\xi_i\in S_m),
\end{equation} 
where $\zeta_m=\frac{1}{n_m}\sum^n_{i=1}\xi_i\cdot I(\xi_i\in S_m)$ is the centroid of cluster $S_m$, with $n_m$ as the number of random variables in $S_m$. 

The total variation distance serves as a homogeneity measure in this clustering step and will also be utilized to determine the borrowing strength in the information borrowing step (see Section \ref{sec:obi} and \ref{sec:embed-obi}). Although other distances such as KL divergence and Hellinger distance can be used as the homogeneity measures, they are equivalent to the total variation distance as they preserve the same order as the total variation in the K-Means algorithm.

\section{Overlapping Indices}
\label{sec:sec3}
The overlapping coefficient (OVL) \citep{Weitzman1970,Schmid2006} is a measure of the intersection area between two probability densities or mass functions. Let $X$ and $Y$ be two random variables with probability densities or mass functions $f$ and $g$, respectively. $\Omega$ is the common support of $f$ and $g$. OVL can be defined using \eqref{eq:ovl-1}, where the integral expression is used in this paper without loss of generality. 
\begin{equation}\label{eq:ovl-1}
     OVL(X,Y) =
     \begin{cases}
        \int_\Omega min\{f(t),g(t)\}\,dt, & \text{continuous distributions}\\
        \sum_\Omega min\{f(t_i),g(t_i)\}I(t_i \in \Omega), & \text{discrete distributions}
    \end{cases}
\end{equation}

An important relationship, $OVL(X,Y)=1-d(X,Y)$, can be obtained from the expressions of OVL \eqref{eq:ovl-1} and the total variation distance \eqref{eq:dtv}. Based on OVL, we define two overlapping indices in the following sections. 

\subsection{Overlapping Clustering Index (OCI)}
To address the challenge of balancing the trade-off between within-cluster homogeneity and efficiency gain through information borrowing, it is necessary to measure the overall within-cluster homogeneity. This can be achieved using the overlapping clustering index (OCI), which is defined as follows:
\begin{definition}[Overlapping Clustering Index]
\label{oci}
Given a set of $n$ random variables with probability density (mass) functions $\{f_1,...,f_n\}$, and partition them into $K$ clusters $S=\{S_1,...,S_K\}$, $1 \le K \le n$. The number of random variables in cluster $S_m$, $m=1,...,K$ is denoted by $n_m = \sum^n_{i=1}I(f_i\in S_m)$ where $I(\cdot)$ is the identity function. The OCI with $K$ clusters partition is defined as follows:
\begin{equation}\label{eq:oci}
     OCI_K=\sum^K_{m=1}p_m\sum^n_{i=1}OVL(g_m,f_i)\cdot I(f_i\in S_m)
\end{equation} 
where $p_m$ denotes the weight with $\sum_{m=1}^K p_m=1$ for the $m$-th cluster, and $g_m$ is the mean of cluster $S_m$, given by:
\begin{equation}\label{eq:g_fun}
     g_m=\frac{1}{n_m}\sum^n_{i=1}f_i \cdot I(f_i\in S_m)
\end{equation}
The weight $p_m$ indicates the preference for the size of cluster $S_m$. We can specify $p_m$ in two settings: (a) $p_m=\frac{1}{K}$, indicating no preference for any cluster, and (b) $p_m = \frac{n_m}{n}$, indicating a preference for big clusters.
\end{definition}

The following propositions discuss the upper bound of $OCI_K$ and its trend as $K$ increases when the set of probability density (mass) functions $\{f_1,...,f_n\}$ is fixed.

\begin{propositionE}\label{prop:oci_interval}
$$OCI_K \in [1, \sum^K_{m=1}p_m n_m]$$
\end{propositionE}

\begin{proofE}
By equation \eqref{eq:g_fun}, we have that
\begin{equation}\label{eq:ovl_detail}
\begin{split}
 OVL(g_m,f_i) & = OVL(f_i,\frac{1}{n_m}\sum^n_{j=1}f_j \cdot I(f_j\in S_m)) \\ 
 & = \int_\Omega min\{f_i(t),\frac{1}{n_m}\sum^n_{j=1}f_j(t) \cdot I(f_j(t)\in S_m)\} \, dt \\
 & = \frac{1}{n_m}\sum^n_{j=1} \int_\Omega min\{f_i(t),f_j(t) \cdot I(f_j(t)\in S_m)\} \, dt \\
 & =\frac{1}{n_m}\sum^n_{j=1}OVL(f_i,f_j) \cdot I(f_j\in S_m).
\end{split}
\end{equation}
Then, by equation \eqref{eq:ovl_detail}, we have that
\begin{equation}\label{eq:oci_k_detail}
\begin{split}
 OCI_K & = \sum^K_{m=1}p_m\sum^n_{i=1}OVL(g_m,f_i)\cdot I(f_i\in S_m) \\ 
 & =\sum^K_{m=1}\frac{p_m}{n_m}\sum^n_{i=1}\sum^n_{j=1}OVL(f_i,f_j) \cdot I(f_j\in S_m)\cdot I(f_i\in S_m) \\
 & = \sum^K_{m=1}\frac{p_m}{n_m}\sum_{f_i,f_j \in S_m}OVL(f_i,f_j).
\end{split}
\end{equation}
Let's consider the following two extreme cases:
\begin{itemize}
    \item Lower bound: If any pair $f_i,f_j$, $i\neq j$ and $i,j \in 1,...,n$, are not overlapped, then $OVL(f_i,f_j)=0$, if $i\neq j$. By \eqref{eq:ovl_detail}, we can calculate that
        $$OCI_K=\sum^K_{m=1}\frac{p_m}{n_m}\sum_{f_i,f_j \in S_m}OVL(f_i,f_j) = \sum^K_{m=1}\frac{p_m}{n_m}n_m=1.$$
    
    \item Upper bound: In cluster $S_m$, $m=1,...,K$, all $f_i$, $i=1,...,n$ and $(f_i\in S_m)$ are fully overlapped ($f_i=f_j$, when $i\neq j$). By \eqref{eq:ovl_detail}, we can calculate that
        $$OCI_K=\sum^K_{m=1}\frac{p_m}{n_m}\sum_{f_i,f_j \in S_m}OVL(f_i,f_j) = \sum^K_{m=1}\frac{p_m}{n_m}n^2_m=\sum^K_{m=1}p_mn_m.$$
\end{itemize}

Since, in all of the other situations, the value of $OCI_K$ must fall in between the above two extreme values, Proposition \ref{prop:oci_interval} holds, $OCI_K \in [1, \sum^K_{m=1}p_m n_m]$.

\end{proofE}

\begin{propositionE}\label{prop:oci_upper}
The upper bound of $OCI_K$ depends on the number of clusters $K$ and can be represented by a function $u_{oci}(K)$. Then, as $K$ increases, $u_{oci}(K)$ monotonically decreases.
\end{propositionE}

\begin{proofE}
From Proposition \ref{prop:oci_interval} we know that $u_{oci}(K)=\sum^K_{m=1}p_m n_m$. In definition \ref{oci}, we specified $p_m=\frac{1}{K}$ or $\frac{n_m}{n}$. So we prove the proposition in such two cases:
\begin{itemize}
    \item[(1)] When $p_m=\frac{1}{K}$, from Proposition \ref{prop:oci_interval}, we know that $$u_{oci}(K)=\sum^K_{m=1}p_m n_m=\frac{1}{K}\sum^K_{m=1}n_m=\frac{n}{K}.$$
    So in case (1), Proposition \ref{prop:oci_upper} holds.
    \item[(2)] When $p_m=\frac{n_m}{n}$, we know that $$u_{oci}(K)=\sum^K_{m=1}p_m n_m=\sum^K_{m=1}\frac{n^2_m}{n}=\frac{1}{n}\sum^K_{m=1}n^2_m.$$
    Without loss of generality, we select an arbitrary cluster $i$ in cluster $\{1,...,K\}$ and split it into two (not empty) clusters $i'$ and $i''$, $n_i=n_{i'}+n_{i''}$. Then,
    \begin{equation*}
    \begin{split}
     u_{oci}(K+1) & =\frac{1}{n}\sum^{K+1}_{m=1}n^2_m
     =\frac{1}{n}[\sum_{m\neq i}n^2_m+(n^2_{i'}+n^2_{i''})] \\
     & < \frac{1}{n}[\sum_{m\neq i}n^2_m+(n_{i'}+n_{i''})^2] 
     = \frac{1}{n}\sum^K_{m=1}n^2_m 
     = u_{oci}(K).
    \end{split}
    \end{equation*}
    So in case (2), Proposition \ref{prop:oci_upper} holds.
\end{itemize}
\end{proofE}

On the one hand, it is easy to see from the definition of $OCI_K$ that maximizing $OCI_K$ when $K$ is fixed means maximizing the weighted sum of OVL, which implies maximizing the  within-cluster homogeneity. On the other hand, propositions \ref{prop:oci_interval} and \ref{prop:oci_upper} demonstrate that the upper bound of $OCI_K$ decreases as the number of clusters $K$ increases, thus imposing a penalty on the number of clusters. As a result, maximizing $OCI_K$ for $K\in\{1,\ldots,n\}$ not only enhances the within-cluster homogeneity but also tends to choose small $K$ and generate larger cluster (i.e., more elements in the cluster) that is more efficient and robust for information borrowing. In Section \ref{sec:oci-km}, we introduce the approach for maximizing $OCI_K$.

To control the degree of penalty on the number of clusters $K$,  we propose the generalized $OCI_K$, which adds a power parameter to the weight $p_m$.
\begin{definition}[Generalized $OCI_K$]
\label{goci}
\begin{equation}\label{eq:goci}
OCI_K=\sum^K_{m=1}p_m^a\sum^n_{i=1}OVL(g_m,f_i)\cdot I(f_i\in S_m)
\end{equation} 
\end{definition}
Although the parameter $a$ can take any positive real number, we restrict its values to the range $(0,1]$ in our study. A larger value of $a$ corresponds to fewer clusters, while a smaller value of $a$ yields more clusters. Appendix \ref{sec:oci-example} provides an example of clustering results along with their corresponding $OCI_K$ values. The impact of the parameter $a$ and guidelines for selecting its value are discussed in detail in Appendix \ref{sec:parameter-effect}. Moreover, Section \ref{sec:paraSelect} provides a comprehensive example that illustrates the process of selecting the optimal value for parameter $a$.

\subsection{Overlapping Borrowing Index (OBI)}\label{sec:obi}
To address the challenge of dynamically quantifying borrowing strength in different clusters, it is necessary to measure the degree of homogeneity within each cluster. This can be accomplished using the overlapping borrowing index (OBI), which is defined as follows:
\begin{definition}[Overlapping Borrowing Index]
\label{obi}
The $OBI$ of cluster $m$, $m=1,\ldots,K$, is defined as follows:
\begin{equation}\label{eq:obi}
     OBI_m=\frac{2}{n_m(n_m-1)}\sum_{i < j, f_i,f_j \in S_m} OVL(f_i,f_j), \hspace{3mm} i,j \in \{1,...,n\}
\end{equation} 
where the cluster $S_m$ and its size $n_m$ are the same as those in the definition of $OCI_K$.
\end{definition}

The value of $OBI$ is in $[0,1]$, which simplifies the assignment of borrowing strength to clusters with varying levels of homogeneity. A larger value of $OBI$ indicates a more homogeneous cluster and a stronger borrowing of information. We present an example in Appendix \ref{sec:BHMOI-example} that illustrates the values of $OBI$ and how they affect dynamic information borrowing.

\section{Bayesian Hierarchical Modeling with Overlapping Indices (BHMOI)}
\label{sec:sec4}
To finally address the issues and challenges discussed in Section \ref{sec:intro}, we incorporate the overlapping indices with model \eqref{eq:BHM-F} to form a novel framework named Bayesian hierarchical modeling with overlapping indices (BHMOI).
\subsection{The Equivalence between Maximizing $OCI_K$ and Weighted K-Means Clustering} \label{sec:oci-km}
The procedure of maximizing $OCI$ can be accomplished in two steps. Firstly, for each $K$ in ${1,\dots,n}$, find the maximum $OCI^*_K=max\{OCI_1,...,OCI_K\}$ and its corresponding clustering result $S^{(K)(oci)*}$. Secondly, find the overall maximum $OCI^* = max\{OCI^*_1,...,OCI^*_n\}$ and its corresponding clustering result $S^{(oci)*}$. The second step is straightforward, so let us focus on the first step. From the definition of generalized $OCI_K$ in \eqref{eq:goci}, we obtain that
\begin{equation}\label{eq:ocis*}
S^{(K)(oci)*} = \argmax_{S^{(K)}} OCI_K = \argmax_{S^{(K)}} \sum^K_{m=1}p_m^a\sum^n_{i=1}OVL(g_m,f_i)\cdot I(f_i\in S^{(K)}_m),
\end{equation}
where $S^{(K)}$ is an arbitrary $K$-partition, $S^{(K)}_m$ is the $m$-th cluster in the $K$-partition, and $S^{(K)(oci)*}$ is the optimal $K$-partition obtained by maximizing $OCI_K$. Let $\{g^*_1,...,g^*_K\}$ be the cluster means of $S^{(K)(oci)*}$. Then the $OCI^*_K$ can be calculated as follows:
$$OCI^*_K = \sum^K_{m=1}p_m^a\sum^n_{i=1}OVL(g^*_m,f_i)\cdot I(f_i\in S^{(K)(oci)*}_m).$$

To address the argmax problem presented in \eqref{eq:ocis*}, we propose a weighted K-Means algorithm defined in \eqref{eq:kmoci-1}. This algorithm extends the K-Means algorithm presented in \eqref{eq:km} by introducing a powered weight term $(p'_m)^b=(1-p_m)^b$ to control the clustering results.
\begin{definition}[Power weighted K-Means]
\label{pwkm}
\begin{equation}\label{eq:kmoci-1}
\begin{split}
S^{(K)(wkm)*} = & \argmin_{S^{(K)}} \sum^K_{m=1}(p'_m)^b\sum^n_{i=1} d(\zeta_m,\xi_i) \cdot I(\xi_i\in S^{(K)}_m) \\
= & \argmin_{S^{(K)}} \sum^K_{m=1}(1-p_m)^b\sum^n_{i=1} (1-OVL(g_m,f_i)) \cdot I(f_i\in S^{(K)}_m)
\end{split}
\end{equation}
where $p_m$, $S^{(K)}$ and $S^{(K)}_m$ are the same as those in equation \eqref{eq:ocis*}, and $S^{(K)(wkm)*}$ is the optimal $K$-partition obtained by the weighted K-Means algorithm. We define the relationship between parameters $a$ and $b$ as $b=c*a^{-1}$, where $a$ is the parameter in the definition \ref{goci} of generalized $OCI_K$, $c\in[a,+ \infty)$ is a constant real number. Since $a \in (0,1]$, $b$ takes values in $[1,+\infty)$. 
\end{definition}
It is worth noting that  the value of $c$, calculated as $c = a * b$, does not need to be explicitly provided. When $p_m=\frac{1}{K}$, the term $(1-p_m)^b$ becomes a constant, resulting in the parameter $b$ having no effect on the clustering result. However, when $p_m=\frac{n_m}{n}$, the value of $b$ does influence the results. Increasing $b$ tends to generate clusters with more extreme sizes, resulting in the presence of very large or very small clusters, in certain cases, lead to a higher number of clusters due to the inclusion of more small clusters. An example illustrating the effect of parameters $a$ and $b$ under the two scenarios of $p_m$ is provided in Appendix \ref{sec:parameter-effect}, where guidelines for determining the optimal values of parameters $a$ and $b$ are also provided.

\begin{theoremE}[Equivalence between maximizing $OCI_K$ and weighted K-Means algorithm]
\label{th:wkm}
Let $S^{(K)(oci)*}$ and $S^{(K)(wkm)*}$ be the optimal solutions of maximizing $OCI_K$ in \eqref{eq:ocis*} and the weighted K-Means algorithm in \eqref{eq:kmoci-1}, respectively. Then, we have $S^{(K)(oci)*}  \equiv S^{(K)(wkm)*}$, where the symbol ``$\equiv$" denotes the equivalence of two partitions.
\end{theoremE}

\begin{proofE}
According the definition \ref{oci}, we consider the proof under two scenarios, $p_m=\frac{1}{K}$ and $p_m=\frac{n_m}{n}$.
\begin{itemize}
    \item[(1)] When $p_m=\frac{1}{K}$, since $K$, $a$ and $b$ are constants, by equation \eqref{eq:kmoci-1} and \eqref{eq:ocis*}, we obtain that
    \begin{equation}\label{th1-11}
    \begin{split}
     S^{(K)(wkm)*} & = \argmin_{S^{(K)}} \sum^K_{m=1}(p'_m)^b\sum^n_{i=1} (1-OVL(g_m,f_i)) \cdot I(f_i\in S^{(K)}_m) \\ 
     & = \argmin_{S^{(K)}} (\frac{K-1}{K})^b \sum^K_{m=1}\sum^n_{i=1} \underline{ (1-OVL(g_m,f_i))} \cdot I(f_i\in S^{(K)}_m) \\
     & = \argmin_{S^{(K)}} \sum^K_{m=1}\sum^n_{i=1} \underline{ (1-OVL(g_m,f_i))} \cdot I(f_i\in S^{(K)}_m),
    \end{split}
    \end{equation}
    and
    \begin{equation}\label{th1-12}
    \begin{split}
     S^{(K)(oci)*} & = \argmax_{S^{(K)}} \sum^K_{m=1}p^a_m\sum^n_{i=1}OVL(g_m,f_i)\cdot I(f_i\in S^{(K)}_m) \\
     & = \argmax_{S^{(K)}} (\frac{1}{K})^a \sum^K_{m=1}\sum^n_{i=1} \underline{OVL(g_m,f_i))} \cdot I(f_i\in S^{(K)}_m) \\
     & = \argmax_{S^{(K)}} \sum^K_{m=1}\sum^n_{i=1} \underline{OVL(g_m,f_i))} \cdot I(f_i\in S^{(K)}_m).
    \end{split}
    \end{equation}
    Because of $OVL(g_m,f_i)$ and $1-OVL(g_m,f_i)$ taking values in $[0,1]$, the underlined parts in equations \eqref{th1-11} and \eqref{th1-12} always change oppositely. This indicates that
    \begin{equation*}
    \begin{split}
      S^{(K)(wkm)*} & = \argmin_{S^{(K)}} \sum^K_{m=1}\sum^n_{i=1} \underline{(1-OVL(g_m,f_i))} \cdot I(f_i\in S^{(K)}_m) \\
     & \Leftrightarrow \argmax_{S^{(K)}} \sum^K_{m=1}\sum^n_{i=1} \underline{OVL(g_m,f_i))} \cdot I(f_i\in S^{(K)}_m) = S^{(K)(oci)*}
    \end{split}
    \end{equation*}
    where $\Leftrightarrow$ denotes the equivalence relationship. Obviously, for any nonnegative values of parameter $a$ and $b$, the two partition results are always equal.
    
    \item[(2)] When $p_m=\frac{n_m}{n}$, by equation \eqref{eq:kmoci-1} and \eqref{eq:ocis*}, we obtain that
    \begin{equation}\label{th1-1}
    \begin{split}
     S^{(K)(wkm)*} & = \argmin_{S^{(K)}} \sum^K_{m=1}(p'_m)^b\sum^n_{i=1} (1-OVL(g_m,f_i)) \cdot I(f_i\in S^{(K)}_m) \\ 
     & = \argmin_{S^{(K)}} \sum^K_{m=1}\sum^n_{i=1} (p'_m)^b \cdot (1-OVL(g_m,f_i)) \cdot I(f_i\in S^{(K)}_m) \\
     & = \argmin_{S^{(K)}} \sum^K_{m=1}\sum^n_{i=1} \underline{ (1-p_m)^b\cdot (1-OVL(g_m,f_i))} \cdot I(f_i\in S^{(K)}_m),
    \end{split}
    \end{equation}
    and
    \begin{equation}\label{th1-2}
    \begin{split}
     S^{(K)(oci)*} & = \argmax_{S^{(K)}} \sum^K_{m=1}p^a_m\sum^n_{i=1}OVL(g_m,f_i)\cdot I(f_i\in S^{(K)}_m) \\
     & = \argmax_{S^{(K)}} \sum^K_{m=1}\sum^n_{i=1} \underline{p^a_m\cdot OVL(g_m,f_i))} \cdot I(f_i\in S^{(K)}_m).
    \end{split}
    \end{equation}
    It is easy to check that all the values of $p_m$, $1-p_m$, $OVL(g_m,f_i)$, and $1-OVL(g_m,f_i)$ take values in $[0,1]$. Since the relationship between parameter $a$ and $b$ is $b=c*a^{-1}$, where $c\in[a,+\infty)$, the underlined parts in equations \eqref{th1-1} and \eqref{th1-2} always change oppositely. This indicates that
    \begin{equation*}
    \begin{split}
      S^{(K)(wkm)*} & = \argmin_{S^{(K)}} \sum^K_{m=1}\sum^n_{i=1} \underline{ (1-p_m)^b\cdot (1-OVL(g_m,f_i))} \cdot I(f_i\in S^{(K)}_m) \\
     & \Leftrightarrow \argmax_{S^{(K)}} \sum^K_{m=1}\sum^n_{i=1} \underline{p^a_m\cdot OVL(g_m,f_i))} \cdot I(f_i\in S^{(K)}_m) = S^{(K)(oci)*}
    \end{split}
    \end{equation*}
    where $\Leftrightarrow$ denotes the equivalence relationship.
\end{itemize}

So, according to the proofs in cases (1) and (2), the following holds,
$$S^{(K)(oci)*}  \equiv S^{(K)(wkm)*}.$$
\end{proofE}

Theorem \ref{th:wkm} states that the weighted K-Means algorithm can be utilized to obtain the optimal clustering result $S^{(K)(oci)*}$.

\subsection{Constructing BHMOI with Embedding $OCI$ and $OBI$ into BHM}\label{sec:embed-obi}
Assuming that the overall maximum $OCI^*$ is achieved at $K$ clusters, i.e., $S^{(oci)*}=S^{(K)(oci)*}$, we can substitute $S^*$ in model \eqref{eq:BHM-F} with $S^{(K)(oci)*}$. Moreover, we can calculate $OBI^{(K)(oci)*}=\{OBI^{(K)(oci)*}_1,\dots,OBI^{(K)(oci)*}_K\}$ based on $S^{(K)(oci)*}$, and thus obtain 
$$p(\boldsymbol{\eta_{jb}}|S^*)=p(\boldsymbol{\eta_{jb}}|S^{(K)(oci)*})=p(\boldsymbol{\eta_{jb}}|OBI^{(K)(oci)*}_j).$$ 
Let $\boldsymbol{\gamma_{jb}}$ represent the hyperparameters of $p(\boldsymbol{\eta_{jb}} | \boldsymbol{\gamma_{jb}})$, which directly control the strength of information borrowing. We can define a function $s(\cdot)$ that maps $OBI^{(K)(oci)*}_j$ to $\boldsymbol{\gamma_{jb}}$, which plays the role of dynamically controlling the borrowing strength in different clusters. Designing an appropriate function $s(\cdot)$ is a non-trivial task that depends on the formula of $p(\boldsymbol{\eta_b} | \boldsymbol{\gamma_{b}})$ and the researcher's preference to borrowing strength (see example in Section \ref{sec:simModel}). By incorporating $s(\cdot)$, the model \eqref{eq:BHM-F} can be rewritten as:
\begin{equation}\label{eq:BHMOI}
     p(\boldsymbol{\theta_j} \mid \boldsymbol{y_j}) \approx \frac {p(\boldsymbol{y_j}\mid \boldsymbol{\theta_j} )}{p(\boldsymbol{y_j})} \int p(\boldsymbol{\theta_j} \mid \boldsymbol{\eta_{jb}},\boldsymbol{\eta_{jo}} ) p(\boldsymbol{\eta_{jo}}) p(\boldsymbol{\eta_{jb}}|s(OBI^{(K)(oci)*}_j)) \,d \boldsymbol{\eta_{jo}} \,d \boldsymbol{\eta_{jb}}
\end{equation}

We refer to equation \eqref{eq:BHMOI} as BHMOI (Bayesian hierarchical model with overlapping indices) which is a flexible framework that can be adapted to different applications. The models constructed using BHMOI involves two steps: (1) applying the K-Means algorithm to maximize OCI and identify the optimal clustering result, and (2) constructing the function $\boldsymbol{\gamma_{jb}}=s(OBI)$ to dynamically set the strength of borrowing. Two models can be found in Section \ref{sec:simModel} as examples. In Appendix \ref{sec:BHMOI-example}, we present a visual example that showcases the capabilities of BHMOI in dynamic information borrowing.

\section{Simulation Study and Real Data Analysis}
\label{sec:sec5}
To address the challenge of analyzing data from multiple subgroups without external data, \cite{Chen2019} proposed a classification-based model called Bayesian hierarchical classification and information sharing (BaCIS). This method involves two steps. In the first step, a classification model is used to classify subgroups into one or two clusters. In the second step, a BHM is employed to borrow information within each cluster. However, BaCIS can only handle problems with two (or one) clusters. To address this limitation, \cite{Chen2020} developed a clustering-based approach, called Bayesian cluster hierarchical model (BCHM), to handle cases with more than two clusters. Similar to BaCIS, BCHM also consists of two steps. In the first step, a Dirichlet process is used to partition subgroups into clusters. In the second step, information borrowing is conducted based on the similarities between subgroups. In this section, we compare BHMOI with these methods. To evaluate their performance, we conduct a simulation study using basket trial data with binary endpoints. We assess the three methods based on two types of operating characteristics: parameter estimation (OCE) and power in hypothesis testing with controlled type 1 error (OCH). Finally, we compare the three methods in a real dataset and discuss their respective results.

\subsection{Model Specification}\label{sec:simModel}
    
\begin{figure}[t!]
\centering
\includegraphics[width=0.4\textwidth]{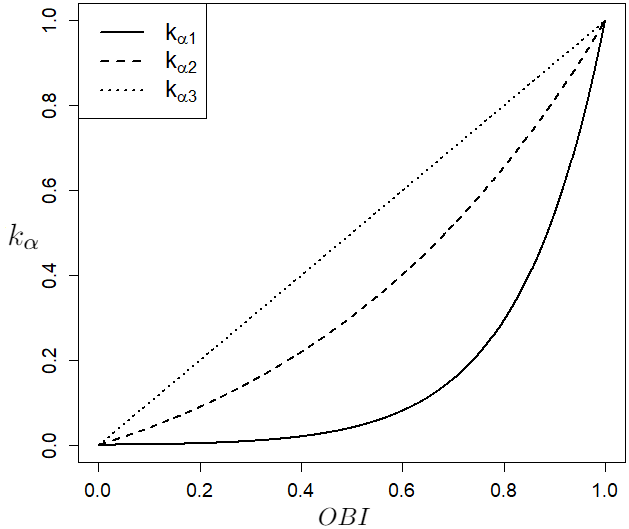}
\caption{The functions of $k_{\alpha}(OBI)$}
\label{fig:s_function}
\end{figure}

Two models are constructed using BHMOI based on the types of endpoints, as shown in Table \ref{tb:2BHMOI}. In both models, the K-Means algorithm is applied to maximize $OCI$ and determine the optimal clustering result $S^{(oci)*}$. The borrowing strength is controlled through a function $\alpha = s(OBI^{(oci)*}_{\boldsymbol{j}}) = \alpha_{min} + k_{\alpha}(OBI^{(oci)*}_{\boldsymbol{j}})(\alpha_{max}-\alpha_{min})$, where $\alpha_{min}$ and $\alpha_{max}$ are pre-specified constant real values that correspond approximately to no borrowing and full borrowing based on the researcher's preference in the application. For this simulation, we set $\beta = 10$, $\alpha_{min}=1$, and $\alpha_{max}=200$. The slope function $k_{\alpha}$ plays a crucial role in determining the borrowing strength based on the value of $OBI$. Figure \ref{fig:s_function} shows different $k_{\alpha}$ functions, $k_{\alpha1} = OBI*e^{-5*(1-OBI)}$, $k_{\alpha2} = OBI*e^{-(1-OBI)}$, and $k_{\alpha3} = OBI$, which reflect researchers' different attitudes toward information borrowing. In this simulation, we choose $k_{\alpha}=k_{\alpha1} = OBIe^{-5*(1-OBI)}$ to ensure that strong borrowing only occurs in the highly homogeneous cluster.

\begin{table*}[t!]
\renewcommand{\arraystretch}{1.5}
\caption{BHMOI for continuous (normal) and binary endpoints}
\centering
\footnotesize
\begin{tabular}{|l|c|c|}
  \hline
  \textbf{Endpoint} & \textbf{Continuous(Normal)} & \textbf{Binary} \\
  \hline \hline
  \multirow{ 2}{*}{\textbf{Model}} & $Y_i \sim N(\theta_i, \frac{1}{\tau_y}), \hspace{3mm} \tau_y \sim gamma(\alpha_y, \beta_y)$ & $Y_i \sim Binomial(N_i, p_i), \hspace{3mm} logit(p_i)=\theta_i$ \\ 
  \cline{2-3}
  & \multicolumn{2}{c|}{
     $\theta_i \sim N(\mu_{\boldsymbol{j}}, \frac{1}{\tau_{\boldsymbol{j}}}), \hspace{3mm} \mu_{\boldsymbol{j}} \sim N(\mu_{0}, \frac{1}{\tau_{0}}), \hspace{3mm} \tau_{\boldsymbol{j}} \sim gamma(\alpha, \beta), \hspace{3mm} \boldsymbol{j}=\{j_1,\dots,j_{n_j}\} \in S^{(oci)*}$
  } \\ 
  \hline
  \hline
  \multirow{ 2}{*}{\textbf{\footnotesize Parameters}} & $\boldsymbol{\eta_{o}} = \{\tau_y,\mu_{\boldsymbol{j}}\}$ & $\boldsymbol{\eta_{o}} = \{\mu_{\boldsymbol{j}}\}$ \\
  \cline{2-3}
  & \multicolumn{2}{c|}{
  $p(\boldsymbol{\eta_{c}}) = p(S^{(oci)*}), \hspace{3mm} \boldsymbol{\eta_{b}} = \{\tau_{\boldsymbol{j}}\}, \hspace{3mm} \boldsymbol{\gamma_b}=\{\alpha,\beta\}, \hspace{3mm} \alpha=s(OBI^{(oci)*}_{\boldsymbol{j}})$
  } \\
  \hline
\end{tabular}
\label{tb:2BHMOI}
\end{table*}

\subsection{Simulation Study}\label{sec:simDesign}

\begin{table*}[t!]
\caption{Different scenarios of heterogeneity in simulation}
\centering
\begin{tabular}{|c|c|c|c|}
  \hline
   & \multicolumn{3}{c|}{Subgroups in each level of response rates} \\ \cline{2-4}
  Scenario & Low ($p=0.1$) & Medium ($p=0.25$) & High ($p=0.5$) \\
  \hline \hline
  1 & $g_1,g_2,g_3,g_4$ & $g_5,g_6,g_7$ & $g_8,g_9,g_{10}$ \\
  2 & $g_1,...,g_7$ & $g_8,g_9$ & $g_{10}$ \\
  3 & $g_1,...,g_5$ & $g_6,...,g_{10}$ & ---  \\
  4 & $g_1,...,g_7$ & --- & $g_8,g_9,g_{10}$ \\
  5 & $g_1,...,g_7$ & $g_8,g_9,g_{10}$ & --- \\
  6 & $g_1,...,g_{10}$ & --- & --- \\
  \hline
\end{tabular}
\label{tb:simDesign}
\end{table*}

To generate simulation data, we propose a basket trial with a total of 150 patients, divided into 10 subgroups with 15 patients in each subgroup. The true response rate ($p$) takes three possible levels: low ($p=0.1$), medium ($p=0.25$), and high ($p=0.5$). The subgroups within each response rate level are exchangeable. To assess the performance of the models in a wide range of situations, we design six scenarios  (see Table \ref{tb:simDesign}) with different combinations of response rate levels, representing various degrees of heterogeneity. We generate 2000 simulated trials for each scenario. Including BHMOI, BaCIS, and BCHM, a total of nine models are compared. Table \ref{tb:simModels} summarizes these models and their settings for information borrowing (except BHMOI). The BHMOI model is the binary endpoint model specified in Table \ref{tb:2BHMOI}. The traditional BHM does not consider heterogeneity among subgroups and assumes that all the subgroups ${g_1,...,g_{10}}$ are exchangeable. The parameter settings of BaCIS and BCHM can be found in the papers \cite{Chen2019} and \cite{Chen2020}.

\subsubsection{Parameter selection}\label{sec:paraSelect}
\begin{figure}[t!]
\centering
\includegraphics[width=\textwidth]{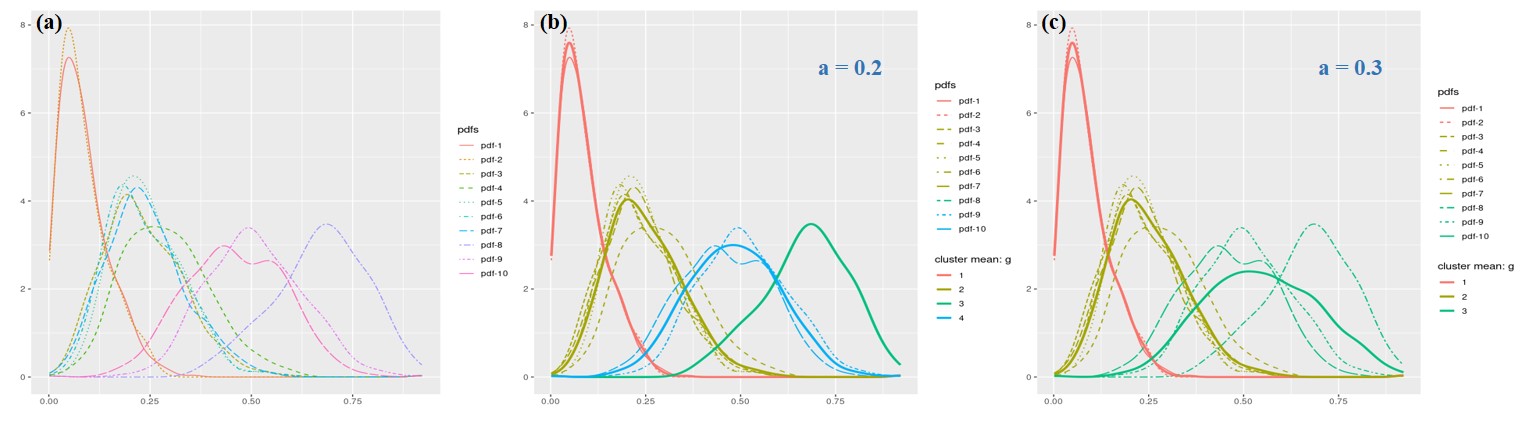}
\caption{Parameter selection}
\label{fig:simParaSelect}
\end{figure}

In this simulation, we set $p_m=\frac{1}{K}$, indicating that we do not have a preference for extreme clusters. In this case, the parameter $b$ does not have any effect on the clustering result. The procedures for selecting parameter $a$ are illustrated in Figure \ref{fig:simParaSelect}. First, we randomly select a data realization from Scenario 1, and the posteriors with non-informative prior are shown in panel (a). We then conduct the weighted K-Means algorithm with different values of parameter $a$, specifically $a=0.2$ and $a=0.3$. The corresponding clustering results are shown in panels (b) and (c), respectively. We prefer the clustering results in panel (c) ($a=0.3$) for two reasons. Firstly, from a robustness perspective, cluster 3 (green) in panel (b) has only one element (subgroup 8), which could make the estimation less robust than in panel (c). Secondly, from a practical perspective, high and very high response rates are usually not distinguished in early-phase trials. In this simulation, we consider $p=0.5$ as the high response rate, so we group subgroups 8, 9, and 10 together, as shown in panel (c). Therefore, we choose parameter $a=0.3$ for Scenario 1. Using the same idea, we select $a=0.3$, $a=0.45$, $a=0.45$, $a=0.45$, and $a=0.5$ for Scenarios 2, 3, 4, 5, and 6, respectively.

\subsubsection{Simulation results}
\begin{figure}[t!]
\centering
\includegraphics[width=.9\textwidth]{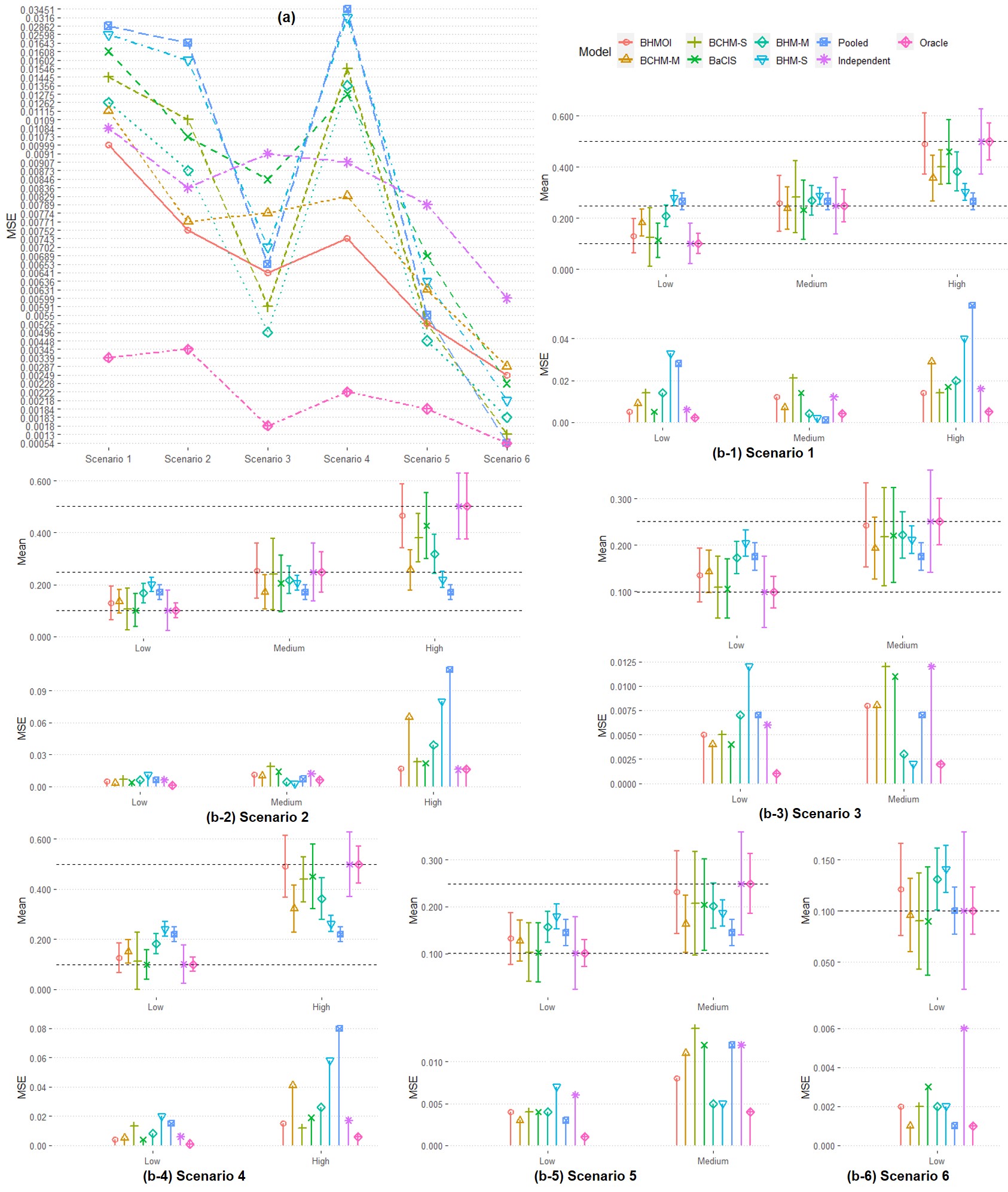}
\caption{The comparison of OCE between BHMOI and existing methods}
\label{fig:sim2msedetails}
\end{figure}

The simulation results are presented in Figure \ref{fig:sim2msedetails} and Table \ref{tb:sim2-pw}. Figure \ref{fig:sim2msedetails} compares the parameter estimation (OCE) performance of the different models. BHMOI outperforms other existing models (excluding the Oracle) in most scenarios, as shown in panel (a). More importantly, BHMOI performs better under heterogeneous circumstances (scenarios 1 to 5) and provides less bias in the estimates. In particular, in heterogeneous scenarios with imbalances (i.e., including both small and large clusters), BHMOI yields less bias and more accurate estimates for small clusters, such as high clusters in scenarios 2 and 4, and medium clusters in scenario 5. In comparison, the existing models, including BaCIS and BCHM, are less accurate and more biased in their estimates.

Table \ref{tb:sim2-pw} presents the comparison of the operating characteristics in hypothesis testing (OCH). We have excluded three models, namely Pooled, BHM-S, and BCHM-S, due to their poor OCH results. BHMOI exhibits greater power in heterogeneous scenarios (scenarios 1 to 5) compared to BCHM-M, BaCIS, and BHM-M, while maintaining comparable Type I errors. Although BHM-M provides more accurate estimates (smaller MSE) than BHMOI in scenarios 3 and 5 (see Figure \ref{fig:sim2msedetails} panel (a)), BHMOI outperforms BHM-M in terms of OCH. This is due to the less biased estimates provided by BHMOI compared to BHM-M (see Figure \ref{fig:sim2msedetails} panels (b-3) and (b-5)), and the unbiasedness of estimation is crucial for good performance in OCH.

\begin{table*}[t!]
\begin{onehalfspace}
\caption{The comparison of OCH between BHMOI and existing methods}
\centering
\scriptsize
\begin{tabularx}{\textwidth}{c *{10}{Y}}
\toprule
Scenarios and methods
 & \multicolumn{10}{c}{Response rate in each subgroup}\\
\cmidrule(lr){1-1} \cmidrule(lr){2-11}
\setrow{\bfseries \itshape} Scenario 1  & 0.1 &  0.1 &  0.1 & 0.1 & 0.25 & 0.25 &  0.25 &  0.5 & 0.5 & 0.5\\
\cmidrule(lr){2-11}
 BHMOI  & 0.1095 & 0.1030 & 0.1000 & 0.0975 & 0.6140 & 0.6005 & 0.5875 & 0.9855 & 0.9900 & 0.9885 \\
 BCHM-M & 0.1130 & 0.1055 & 0.1185 & 0.1085 & 0.5265 & 0.5325 & 0.5605 & 0.9670 & 0.9730 & 0.9765 \\
 BaCIS  & 0.2025 & 0.2055 & 0.1975 & 0.2095 & 0.7185 & 0.6825 & 0.6905 & 0.9845 & 0.9890 & 0.9865\\
 BHM-M  & 0.1260 & 0.1335 & 0.1130 & 0.1190 & 0.5270 & 0.5180 & 0.5265 & 0.9515 & 0.9565 & 0.9610\\
 Independent  & 0.0620 & 0.0610 & 0.0565 & 0.0545 & 0.5400 & 0.5335 & 0.5300 & 0.9830 & 0.9845 & 0.9865\\
 Oracle  & 0.0070 & 0.0070 & 0.0070 & 0.0070 & 0.7070 & 0.7070 & 0.7070 & 1.0000 & 1.0000 & 1.0000\\
\cmidrule(lr){2-11}
\setrow{\bfseries \itshape} Scenario 2  & 0.1 &  0.1 &  0.1 & 0.1 & 0.1 & 0.1 &  0.1 &  0.25 &  0.25 & 0.5\\
\cmidrule(lr){2-11}
 BHMOI  & 0.0895 & 0.0825 & 0.0865 & 0.0865 & 0.0790 & 0.0730 & 0.0825 & 0.5515 & 0.5455 & 0.9830 \\
 BCHM-M & 0.0875 & 0.0840 & 0.0890 & 0.0950 & 0.0635 & 0.0775 & 0.0890 & 0.4250 & 0.4250 & 0.9380 \\
 BaCIS  & 0.1020 & 0.1155 & 0.1245 & 0.1195 & 0.1065 & 0.0980 & 0.1095 & 0.5645 & 0.5450 & 0.9245\\
 BHM-M  & 0.0925 & 0.0895 & 0.1095 & 0.0990 & 0.0915 & 0.0965 & 0.0970 & 0.4910 & 0.4875 & 0.9470\\
 Independent  & 0.0605 & 0.0540 & 0.0595 & 0.0655 & 0.0550 & 0.0485 & 0.0565 & 0.5425 & 0.5245 & 0.9860\\
 Oracle  & 0.0010 & 0.0010 & 0.0010 & 0.0010 & 0.0010 & 0.0010 & 0.0010 & 0.6445 & 0.6445 & 0.9860\\
\cmidrule(lr){2-11}
\setrow{\bfseries \itshape} Scenario 3  & 0.1 &  0.1 &  0.1 & 0.1 & 0.1 & 0.25 &  0.25 &  0.25 & 0.25 & 0.25\\
\cmidrule(lr){2-11}
 BHMOI  & 0.0960 & 0.1030 & 0.0950 & 0.0945 & 0.0980 & 0.5750 & 0.6120 & 0.5835 & 0.5990 & 0.5980 \\
 BCHM-M & 0.1000 & 0.1185 & 0.1090 & 0.1190 & 0.1140 & 0.5605 & 0.6060 & 0.5675 & 0.5815 & 0.5940 \\
 BaCIS  & 0.0945 & 0.0980 & 0.1060 & 0.1025 & 0.1115 & 0.5010 & 0.5190 & 0.5235 & 0.5195 & 0.5195\\
 BHM-M  & 0.1315 & 0.1365 & 0.1270 & 0.1275 & 0.1295 & 0.5520 & 0.5675 & 0.5595 & 0.5595 & 0.5520\\
 Independent  & 0.0515 & 0.0575 & 0.0515 & 0.0590 & 0.0635 & 0.5330 & 0.5500 & 0.5400 & 0.5520 & 0.5485\\
 Oracle  & 0.0015 & 0.0015 & 0.0015 & 0.0015 & 0.0015 & 0.8140 & 0.8140 & 0.8140 & 0.8140 &     0.8140\\
\cmidrule(lr){2-11}
\setrow{\bfseries \itshape} Scenario 4  & 0.1 &  0.1 &  0.1 & 0.1 & 0.1 &  0.1 & 0.1 &  0.5 & 0.5 & 0.5\\
\cmidrule(lr){2-11}
 BHMOI  & 0.0580 & 0.0410 & 0.0485 & 0.0560 & 0.0540 & 0.0485 & 0.0505 & 0.9760 & 0.9720 & 0.9680 \\
 BCHM-M & 0.0520 & 0.0440 & 0.0605 & 0.0545 & 0.0545 & 0.0505 & 0.0555 & 0.9230 & 0.9375 & 0.9340 \\
 BaCIS  & 0.1470 & 0.1460 & 0.1605 & 0.1565 & 0.1485 & 0.1480 & 0.1560 & 0.9805 & 0.9775 & 0.9780\\
 BHM-M  & 0.0760 & 0.0830 & 0.0915 & 0.0855 & 0.0875 & 0.0885 & 0.0910 & 0.9510 & 0.9550 & 0.9550\\
 Independent  & 0.0520 & 0.0455 & 0.0525 & 0.0520 & 0.0575 & 0.0470 & 0.0495 & 0.9785 & 0.9795 & 0.9780\\
 Oracle  & 0.0005 & 0.0005 & 0.0005 & 0.0005 & 0.0005 & 0.0005 & 0.0005 & 1.0000 & 1.0000 & 1.0000\\
\cmidrule(lr){2-11}
\setrow{\bfseries \itshape} Scenario 5  & 0.1 &  0.1 &  0.1 & 0.1 & 0.1 &  0.1 & 0.1 &  0.25 & 0.25 & 0.25\\
\cmidrule(lr){2-11}
 BHMOI  & 0.0820 & 0.0735 & 0.0755 & 0.0890 & 0.0780 & 0.0900 & 0.0845 & 0.5375 & 0.5425 & 0.5390 \\
 BCHM-M & 0.0910 & 0.0875 & 0.0925 & 0.1060 & 0.0870 & 0.1010 & 0.1145 & 0.4400 & 0.4805 & 0.5175 \\
 BaCIS  & 0.0720 & 0.0660 & 0.0615 & 0.0760 & 0.0645 & 0.0635 & 0.0650 & 0.3895 & 0.4005 & 0.3920\\
 BHM-M  & 0.0935 & 0.0890 & 0.0805 & 0.0995 & 0.1010 & 0.1015 & 0.1000 & 0.4685 & 0.4725 & 0.4660\\
 Independent  & 0.0585 & 0.0555 & 0.0560 & 0.0590 & 0.0580 & 0.0615 & 0.0555 & 0.5425 & 0.5320 & 0.5285\\
 Oracle  & 0.0000 & 0.0000 & 0.0000 & 0.0000 & 0.0000 & 0.0000 & 0.0000 & 0.7155 & 0.7155 & 0.7155\\
\cmidrule(lr){2-11}
\setrow{\bfseries \itshape} Scenario 6  & 0.1 &  0.1 &  0.1 & 0.1 & 0.1 &  0.1 & 0.1 &  0.1 & 0.1 & 0.1\\
\cmidrule(lr){2-11}
 BHMOI  & 0.0385 & 0.0300 & 0.0305 & 0.0335 & 0.0315 & 0.0305 & 0.0310 & 0.0365 & 0.0315 & 0.0350 \\
 BCHM-M & 0.0365 & 0.0335 & 0.0395 & 0.0440 & 0.0310 & 0.0245 & 0.0405 & 0.0290 & 0.0310 & 0.0415 \\
 BaCIS  & 0.0150 & 0.0085 & 0.0080 & 0.0090 & 0.0095 & 0.0075 & 0.0105 & 0.0140 & 0.0115 & 0.0115\\
 BHM-M  & 0.0185 & 0.0115 & 0.0125 & 0.0140 & 0.0200 & 0.0145 & 0.0130 & 0.0135 & 0.0150 & 0.0120\\
 Independent  & 0.0625 & 0.0495 & 0.0505 & 0.0555 & 0.0605 & 0.0515 & 0.0535 & 0.0610 & 0.0560 & 0.0545\\
 Oracle  & 0.0000 & 0.0000 & 0.0000 & 0.0000 & 0.0000 & 0.0000 & 0.0000 & 0.0000 & 0.0000 & 0.0000\\
\bottomrule
\end{tabularx}
\label{tb:sim2-pw}
\end{onehalfspace}
\end{table*}

In the previous simulation, we assumed that the response rates of subgroups were the same within each cluster. However, in reality, the response rates of subgroups may differ, and we can generate them from a distribution to maintain exchangeability within clusters. For instance, in the case of three fixed levels used in the simulation, we can generate response rates for subgroups from three uniform distributions with intervals [0.05, 0.15], [0.20, 0.30], and [0.35, 0.65]. Figure \ref{fig:simRandom} in Appendix \ref{apdxB} presents the OCE comparison for this example. The results are similar to those of the previous simulation. As the response rates within each cluster are different, we do not present the OCH table for this example.

\subsection{Real Data Analysis}\label{sec:realExample}

The real data used in this study was collected from a phase II trial conducted by \cite{Chugh2009} to evaluate the efficacy of imatinib in 10 subtypes of advanced sarcoma. These subtypes correspond to 10 subgroups, namely angiosarcoma, Ewing sarcoma, fibrosarcoma, leiomyosarcoma, liposarcoma, malignant fibrous histiocytoma, osteosarcoma, malignant peripheral nerve sheath tumor, rhabdomyosarcoma, and synovial subtypes. The number of responses and total number of patients in each subgroup were 2/15, 0/3, 1/12, 6/28, 7/29, 3/29, 5/26, 1/5, 0/2, and 3/20, respectively. These observed response rates range widely from 0\% to 24.1\%. 

\begin{figure}[t!]
\centering
\includegraphics[width=\textwidth]{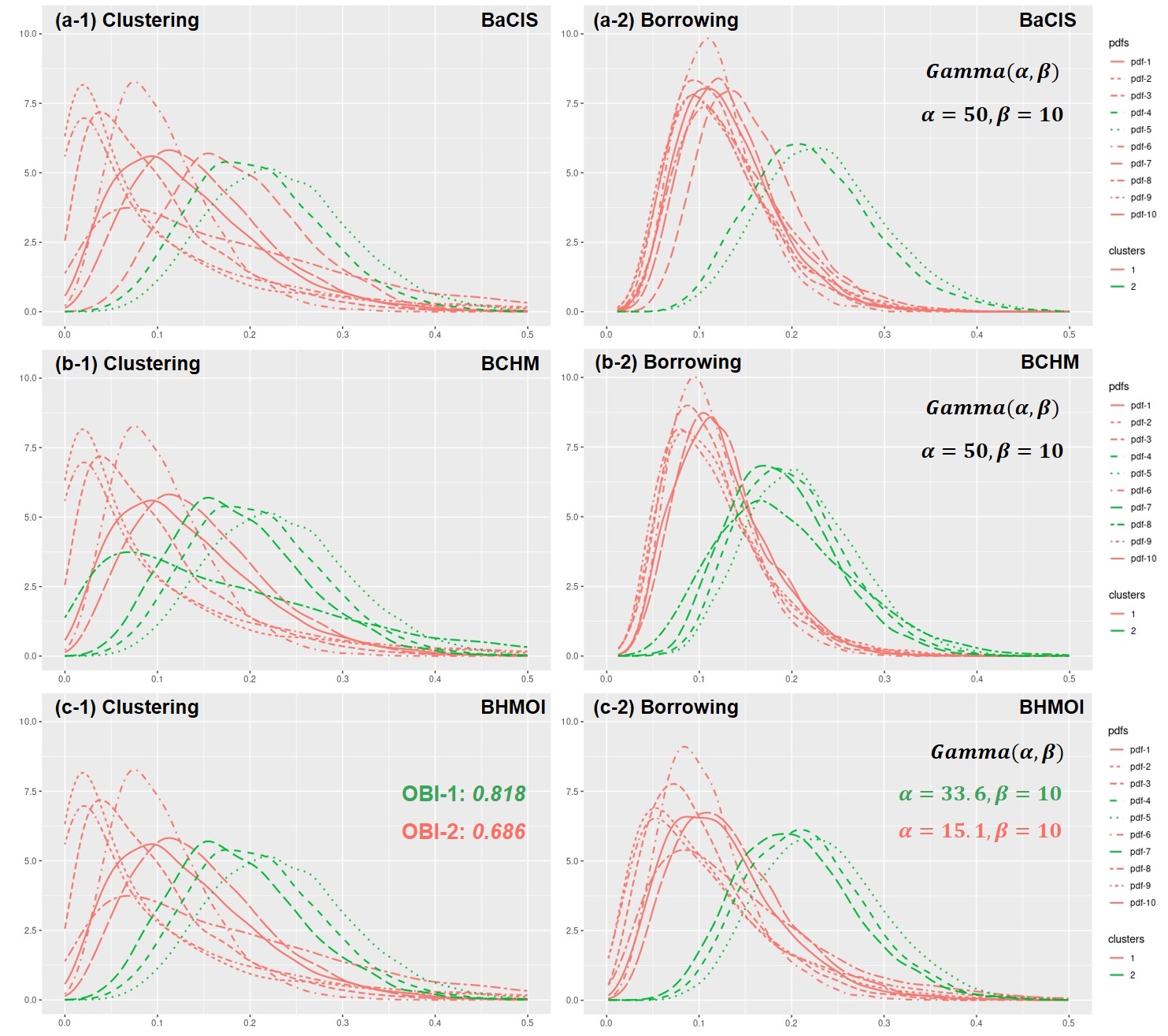}
\caption{The comparison of BHMOI, BCHM and BaCIS on real data example}
\label{fig:real_compare}
\end{figure}

\cite{Chen2020} compared the results of BCHM, BaCIS, and traditional BHM in their analysis. The traditional hierarchical model assumes all subgroups belong to the same pool, which can significantly lower the posterior response rate of subgroups with low response rates. BCHM, on the other hand, identified two clusters in the dataset (the left column of Figure \ref{fig:real_compare} displays the clusters of posteriors with non-informative prior), with a low response rate cluster including subgroups 1, 2, 3, 6, 9, and 10, and a high response rate cluster including subgroups 4, 5, 7, and 8, as shown in panel (b-1) of Figure \ref{fig:real_compare}. In comparison, BaCIS only classified subgroups 4 and 5 into the high response cluster, while subgroups 7 and 8 were classified into the low response cluster, as shown in panel (a-1) of Figure \ref{fig:real_compare}. 

\begin{figure}[t!]
\centering
\includegraphics[width=\textwidth]{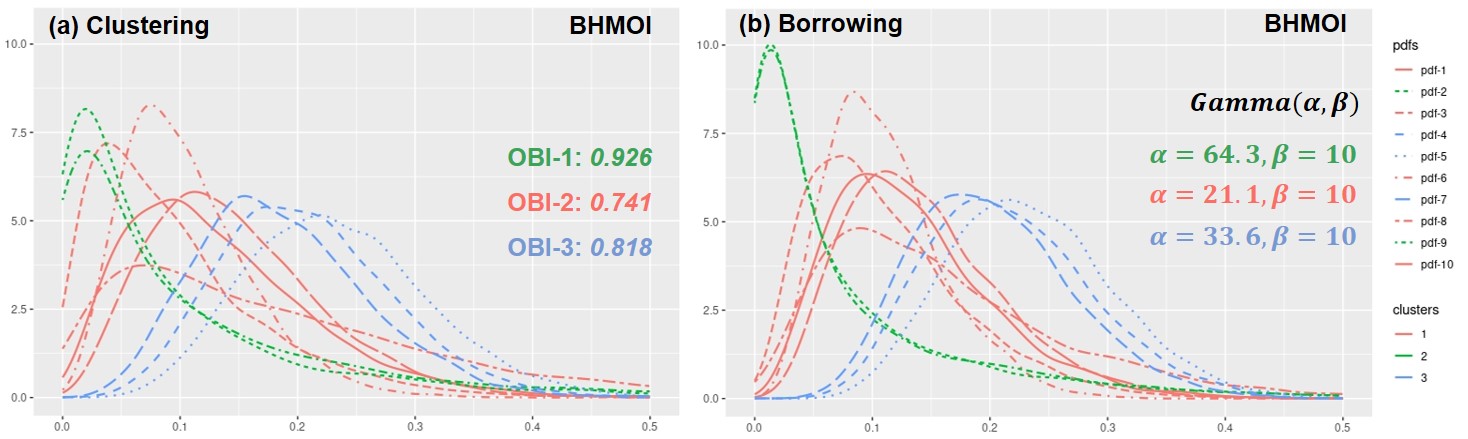}
\caption{The BHMOI model with 3 clusters}
\label{fig:real_3c}
\end{figure}

For BHMOI, we set $\beta=10$, $\alpha = s(OBI)= \alpha_{min} + OBI*e^{-5*(1-OBI)}(\alpha_{max}-\alpha_{min})$, where $\alpha_{min}=1$ and $\alpha_{max}=100$ correspond to no and full information borrowing, respectively. By choosing parameter $a=0.25$, BHMOI identifies two clusters, where subgroups 4, 5, and 7 belong to the high cluster and the remaining subgroups belong to the low cluster. The plot of posteriors with non-informative prior indicates that the clustering result of BHMOI is more reasonable compared to those of BCHM and BaCIS. Furthermore, the borrowing strength of each cluster, controlled by the parameter $\alpha$ in the prior $Gamma(\alpha,\beta)$, is dynamically determined by the value of $OBI$ of each cluster (see panel (c-2) of Figure \ref{fig:real_compare}). The low cluster, being more heterogeneous, has less information borrowing than the high cluster, which is more homogeneous. When we set parameter $a=0.2$, indicating a preference for more homogeneous clusters, BHMOI identifies three clusters with higher average $OBI$, as shown in panel (a) of Figure \ref{fig:real_3c}. The borrowing strength of clusters are specified in panel (b) of Figure \ref{fig:real_3c}.

\section{Discussion}
\label{sec:Con}

We have introduced a BHM framework along with two innovative overlapping indices, OCI and OBI. By integrating these indices into the framework, we have developed a method called BHMOI, which facilitates heterogeneity identification and dynamic information borrowing in subgroup data analysis. Simulation and real data studies have demonstrated that BHMOI outperforms existing methods, exhibiting improved operating characteristics in both estimation and hypothesis testing (OCE and OCH), particularly in unbalanced heterogeneous scenarios. BHMOI has a wide range of potential applications, such as providing a flexible, efficient, and robust analytical platform for medical research in synthesizing information from experiments and clinical trials across different experimental units or patient subgroups, which naturally occurs in master protocol trial designs such as basket, umbrella, and platform trials \citep{Chen2019, Chen2020, Hobbs2018, Chu2018, Yin2021, Lu_2021, Berry2020}. It can also assist better decision-making in many areas, including the design of future experiments, drug development, and the optimization of patient treatment strategies in precision medicine. Our current work did not account for the potential impact of covariates, which could offer valuable additional information. Exploring the effects of covariates is a promising avenue for future research.

\newpage
\newpage
\bibliographystyle{agsm}

\bibliography{docbi}

\newpage

\noindent\textbf{\large Acknowledgments}

The authors appreciate the editorial assistance from Mrs. Jessica Swann.

\noindent\textbf{\large Funding}

J. Jack Lee's research was supported in part by the grants P30CA016672, P50CA221703, U24CA224285, and U24CA274274 from the National Cancer Institute.

\newpage

\appendix
\counterwithin{figure}{section}
\counterwithin{table}{section}
\begin{appendices}

\section{Proofs and Examples}\label{apdxA}
\subsection{Proofs}\label{apdxAproof}
\printProofs

\subsection{Clustering examples and effects of parameters}\label{apdxAexample}
\subsubsection{Clustering examples}\label{sec:oci-example}

\begin{figure}[t!]
\centering
\includegraphics[width=\textwidth]{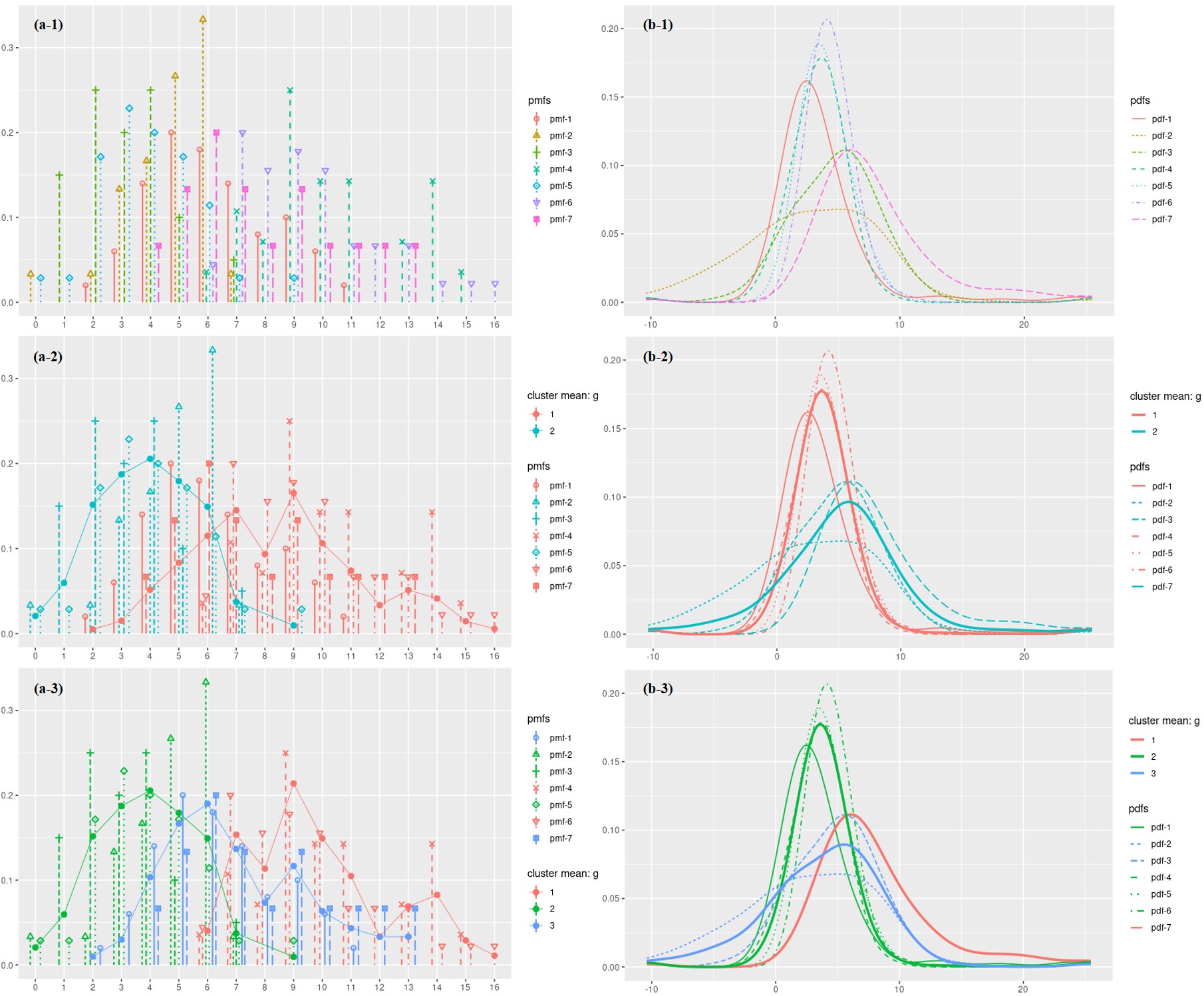}
\caption{Examples of clustering discrete (left) and continuous (right) distributions}
\label{fig:examples}
\end{figure}

Figure \ref{fig:examples} presents two examples of clustering discrete and continuous distributions. The left panels (a-1 to a-3) show the clustering of discrete distributions, while the right panels (b-1 to b-3) show the clustering of continuous distributions. The distributions to be clustered are shown in the first row. We set the number of clusters to $K=2$ and $K=3$, and the clustering results are presented in the second and third rows, respectively. The means of each cluster are also shown. The values of $OCI_K$ can be calculated using equation \eqref{eq:goci} with $a=1$. For the discrete example, the values of $OCI_K$ are $OCI_2=2.498$ and $OCI_3=1.907$, while for the continuous example, the values of $OCI_K$ are $OCI_2=2.464$ and $OCI_3=1.797$. Based on these results, it is clear that when maximizing $OCI_K$, two clusters are preferable to three clusters in both examples.

\subsubsection{Effects of parameters}\label{sec:parameter-effect}
The weighted K-Means algorithm is advantageous due to its simplicity, relying on only two parameters: $a$ in \eqref{eq:goci} (Definition \ref{goci}, generalized OCI) and $b$ in \eqref{eq:kmoci-1} (Definition \ref{pwkm}, power-weighted K-Means). The proof of Theorem \ref{th:wkm} illustrates that when $p_m=\frac{1}{K}$, both $a$ and $b$ have no impact on the optimal $K$ clustering results. However, increasing $a$ influences all values of ${OCI_1, ..., OCI_n}$ and determines the overall optimal clustering result by reducing the number of clusters. When $p_m=\frac{n_m}{n}$, both $a$ and $b$ influence the optimal $K$ clustering result. Notably, $b$ can be expressed as $b=c*a^{-1}$. A large (small) value of $b$ ($a$) tends to generate more extreme clusters, characterized by significantly large or small sizes. Similar to the scenario with $p_m=\frac{1}{K}$, parameter $a$ also affects the values of ${OCI_1, ..., OCI_n}$ and determines the overall optimal clustering result. To illustrate the effects of parameters $a$ and $b$, we provide an example employing a BHM with the following specifications:
$$y_i \sim Bin(n, p_i), \hspace{3mm} logit(p_i)=\theta_i, \hspace{3mm} \theta_i \sim N(\mu, \frac{1}{\tau}).$$
where $n=25$ and $\mu=logit(0.1)$. We set $\tau=0.01$ to indicate that we provide a noninformative prior for $\theta_i$. The observed data are $y_1=2$, $y_2=3$, $y_3=5$, $y_4=6$, $y_5=8$, $y_6=6$, $y_7=9$, $y_8=11$, $y_9=13$, and $y_{10}=7$.

\begin{figure}[t!]
\centering
\includegraphics[width=\textwidth]{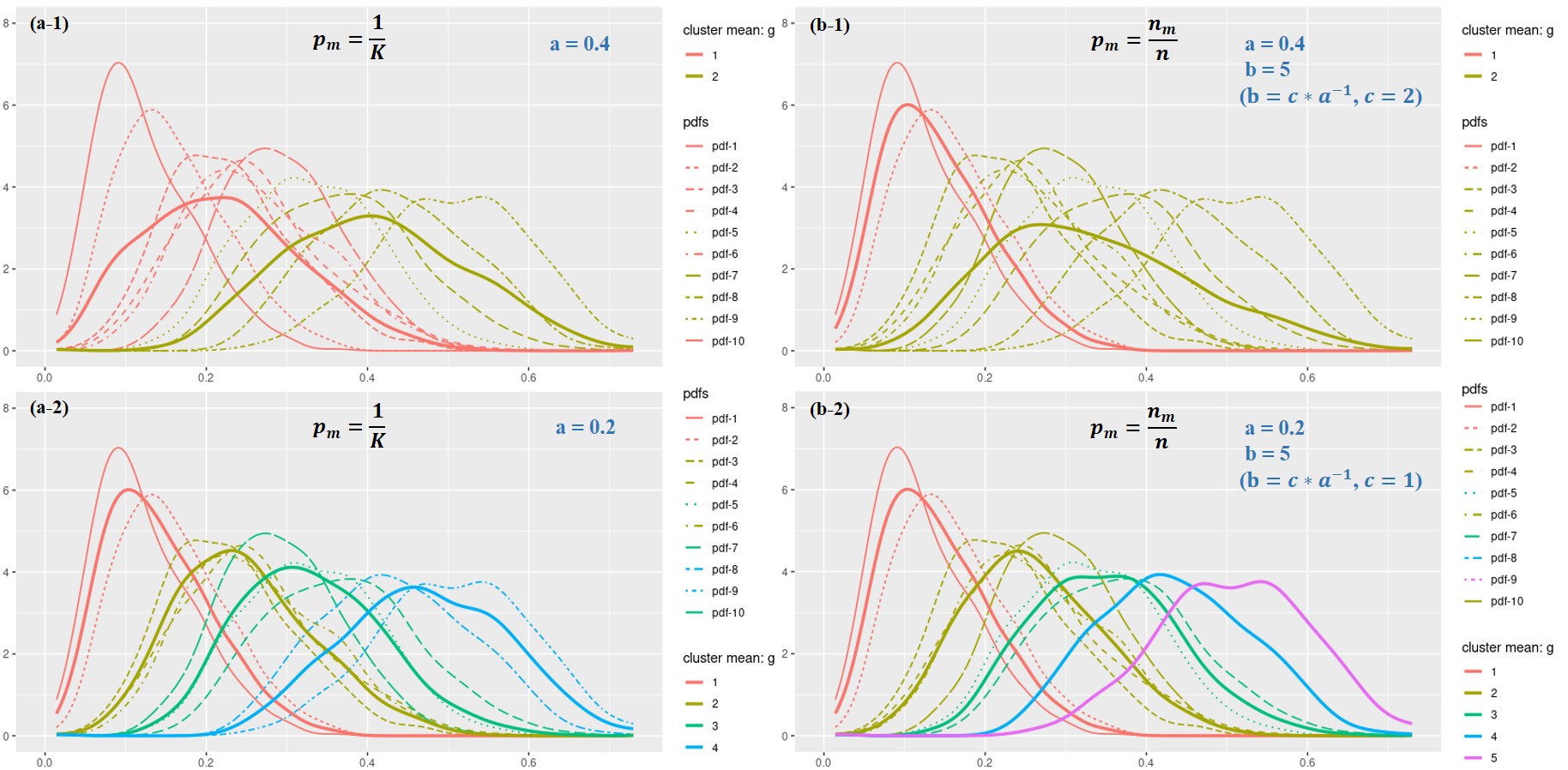}
\caption{The effects of parameter $a$ and $b$ on clustering}
\label{fig:parameters}
\end{figure}

To visually demonstrate the impact of parameters $a$ and $b$ in the scenarios $p_m=\frac{1}{K}$ and $p_m=\frac{n_m}{n}$, we present the optimal clustering results in Figure \ref{fig:parameters}. For the effects of $a$, we compare panels (a-1, b-1) in the first row with panels (a-2, b-2) in the second row of Figure \ref{fig:parameters}. It is observed that as $a$ increases, the number of clusters decreases. To observe the effects of $b$ in the scenario $p_m=\frac{n_m}{n}$, we compare panels (a-1, a-2) in the first column with panels (b-1, b-2) in the second column. Clearly, parameter $b$ has the ability to generate extreme clusters, characterized by significantly large or small sizes.

This example also provides a practical approach to determining optimal parameter values for an application. There are general guidelines for identifying these values:

\begin{itemize}
\item[(1)] Choose $p_m = \frac{1}{K}$ or $p_m = \frac{n_m}{n}$ according the application.
\item[(2)] In the scenario $p_m = \frac{1}{K}$, only parameter $a$ needs to be determined. We can select its value intuitively. By plotting distributions with different values of $a$ using the application data (as shown in the left panels of Figure \ref{fig:parameters}), we can visually assess the clustering results. Based on the refinement nature of the clustering model, we can choose candidate values for $a$ that seem suitable for the application. For instance, from Figure \ref{fig:parameters}, we may consider $a={0.2, 0.35, 0.41}$ as candidate values.
\item[(3)] Conduct simulations (sensitivity analysis) on the candidate values to identify the optimal value.
\item[(4)] In the scenario $p_m = \frac{n_m}{n}$, both parameters $a$ and $b$ need to be determined. We can first follow steps (2) and (3) to find the optimal value for $a$, and then repeat the process to determine the optimal value for $b$.
\end{itemize}

\subsection{An example of dynamic information borrowing using BHMOI}\label{sec:BHMOI-example}
\begin{figure}[t!]
\centering
\includegraphics[width=\textwidth]{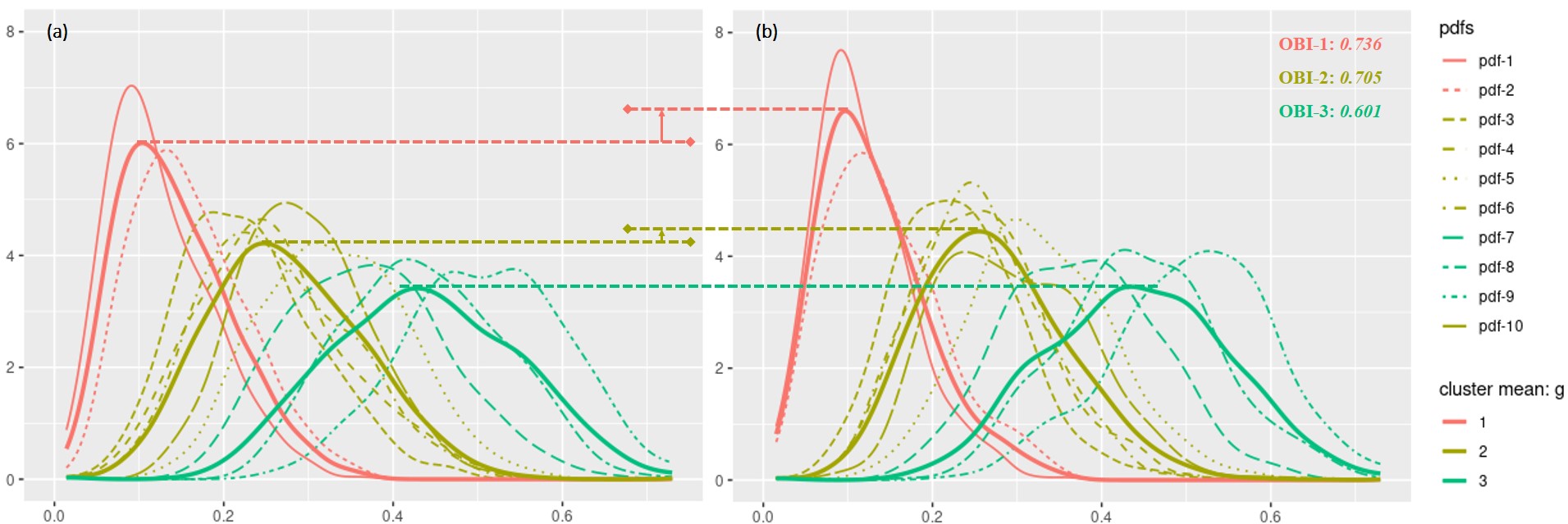}
\caption{Example of dynamic information borrowing based on OBIs}
\label{fig:dynamicBorrowing}
\end{figure}

To visually demonstrate the dynamic information borrowing capability of BHMOI, we use the example in Appendix \ref{sec:parameter-effect} that includes data, a model, and parameter settings. The construction of function $s(\cdot)$ can be found in Section \ref{sec:simModel}. Figure \ref{fig:dynamicBorrowing} depicts the results before (panel (a)) and after (panel (b)) information borrowing. The clustering yields three clusters with OBIs of 0.736, 0.705, and 0.601, respectively. The function $s(\cdot)$ is employed to determine the borrowing strength in each cluster based on these OBIs. Notably, the strength of information borrowing is ordered as follows: cluster 1 (red) $>$ cluster 2 (khaki) $>$ cluster 3 (green), which aligns with the OBI rank.

\newpage

\section{Supplementary Figures and Tables}\label{apdxB}

\begin{table*}[ht]
\renewcommand{\arraystretch}{1.5}
\caption{Models and short descriptions}
\small
\centering
\begin{tabularx}{\textwidth}{|l|X|}
  \hline
  Model & \multicolumn{1}{c|}{Short descriptions}  \\
  \hline \hline
  Oracle & Assumes that we know the true clustering structure and can correctly classify subgroups into the correct clusters. Then, we treat the clusters independently. In each cluster, we do full borrowing, i.e., pool all subgroups together. \\
  \hline
  Independent & No information borrowing; treat each group independently. \\
  \hline
  Pooled & Full borrowing, pool all subgroups together. \\
  \hline
  BHM-M & Traditional BHM with moderate borrowing (borrowing prior: $Gamma(5,1)$). \\
  \hline
  BHM-S & Traditional BHM with strong borrowing (borrowing prior: $Gamma(50,1)$). \\
  \hline
  BaCIS & classification-based model (borrowing prior: $Gamma(50,2)$). \\
  \hline
  BCHM-M & BCHM with moderate borrowing (borrowing prior: $Gamma(50,1)$). \\
  \hline
  BCHM-S & BCHM with strong borrowing (borrowing prior: $Gamma(500,1)$). \\
  \hline
\end{tabularx}
\label{tb:simModels}
\end{table*}

\begin{figure}[ht]
\centering
\includegraphics[width=.88\textwidth]{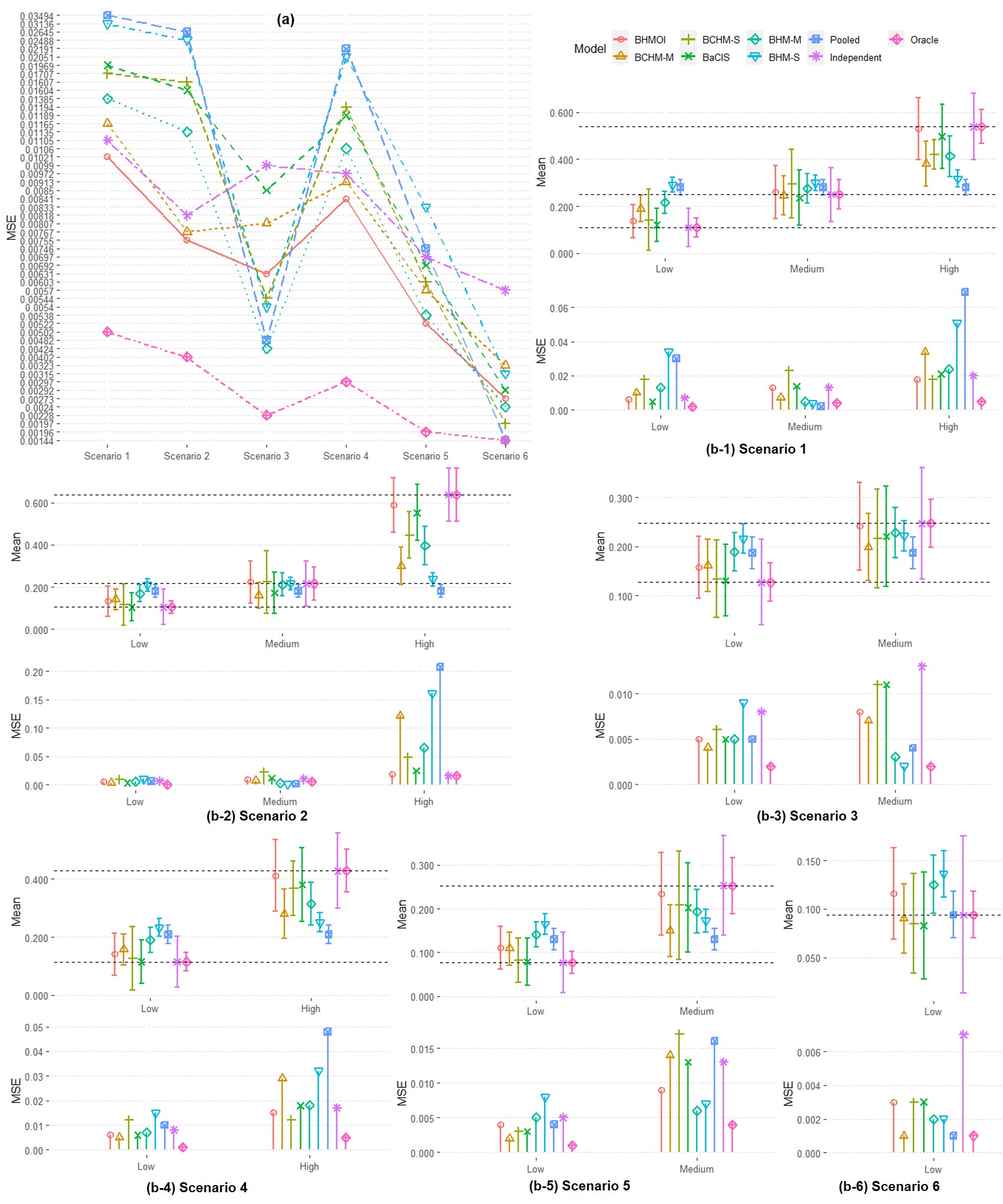}
\caption{Comparisons of OCE between BHMOI and existing methods in a scenario where the response rates of subgroups were generated from uniform distributions.}
\label{fig:simRandom}
\end{figure}

\FloatBarrier

\end{appendices}


\end{document}